\documentclass[prx, letterpaper, superscriptaddress, amsmath, amssymb, amssymb, reprint, floatfix]{revtex4-2}

\usepackage{graphicx}
\usepackage{hyperref}
\usepackage{dcolumn}% Align table columns on decimal point
\usepackage{verbatim}
\usepackage{xr}
\usepackage{mathtools}
\usepackage{physics}
\usepackage{xcolor}

\usepackage{booktabs}
\usepackage{float}

\begin{document}

%\title{Asymmetric Presence of Pauli Spin Blockade and Fast Decoherence in Silicon Nanowire Quantum Dots}
%\title{Energy Level-Selective Pauli Spin Blockade and Fast Decoherence in Silicon Nanowire Quantum Dots}
\title{Non-reciprocal Pauli Spin Blockade in a Silicon Double Quantum Dot}

\author{Theodor Lundberg} \email{twl28@cam.ac.uk}
\affiliation{Cavendish Laboratory, University of Cambridge, J.J. Thomson Avenue, Cambridge CB3 0HE, UK}
\affiliation{Hitachi Cambridge Laboratory, J.J. Thomson Avenue, Cambridge CB3 0HE, UK}
\author{David J. Ibberson} \email{Current address: Quantum Motion Technologies, Windsor House, Cornwall Road, Harrogate HG1 2PW, UK}
\affiliation{Quantum Engineering Technology Labs, University of Bristol, Tyndall Avenue, Bristol BS8 1FD, UK}
\affiliation{Hitachi Cambridge Laboratory, J.J. Thomson Avenue, Cambridge CB3 0HE, UK}
\affiliation{Quantum Engineering Centre for Doctoral Training, University of Bristol, Tyndall Avenue, Bristol BS8 1FD, UK}
\author{Jing Li}
\affiliation{Universit\'{e} Grenoble Alpes, CEA, IRIG, MEM/L\_Sim, 38000 Grenoble, France}
\affiliation{Universit\'{e} Grenoble Alpes, CEA, LETI, 38000 Grenoble, France}
\author{Louis Hutin}
\affiliation{Universit\'{e} Grenoble Alpes, CEA, LETI, 38000 Grenoble, France}
\author{José C. Abadillo-Uriel}
\affiliation{Universit\'{e} Grenoble Alpes, CEA, IRIG, MEM/L\_Sim, 38000 Grenoble, France}
\author{Michele Filippone}
\affiliation{Universit\'{e} Grenoble Alpes, CEA, IRIG, MEM/L\_Sim, 38000 Grenoble, France}
\author{Benoit Bertrand}
\affiliation{Universit\'{e} Grenoble Alpes, CEA, LETI, 38000 Grenoble, France}
\author{Andreas Nunnenkamp}
\affiliation{Faculty of Physics, University of Vienna, Boltzmanngasse 5, 1090 Vienna, Austria}
\author{Chang-Min Lee}
\affiliation{Department of Materials Science and Metallurgy, University of Cambridge, 27 Charles Babbage Road, Cambridge CB3 0FS, UK}
\author{Nadia Stelmashenko}
\affiliation{Department of Materials Science and Metallurgy, University of Cambridge, 27 Charles Babbage Road, Cambridge CB3 0FS, UK}
\author{Jason W. A. Robinson}
\affiliation{Department of Materials Science and Metallurgy, University of Cambridge, 27 Charles Babbage Road, Cambridge CB3 0FS, UK}
\author{Maud Vinet}
\affiliation{Universit\'{e} Grenoble Alpes, CEA, LETI, 38000 Grenoble, France}
\author{Lisa Ibberson}
\affiliation{Hitachi Cambridge Laboratory, J.J. Thomson Avenue, Cambridge CB3 0HE, UK}
\author{Yann-Michel Niquet}
\affiliation{Universit\'{e} Grenoble Alpes, CEA, IRIG, MEM/L\_Sim, 38000 Grenoble, France}
\author{M. Fernando Gonzalez-Zalba} \email{mg507@cam.ac.uk}
\affiliation{Quantum Motion Technologies, Windsor House, Cornwall Road, Harrogate HG1 2PW, UK}

\begin{abstract}
% Spin qubits in gate-defined silicon quantum dots are receiving increased attention and being fabricated in industrial cleanrooms thanks to their potential for large-scale quantum computing. 

%Readout of silicon quantum dot spin qubits is done most accurately and scalably via Pauli spin blockade. Various mechanisms may however lift the blockade and complicate readout. In this work, we expand the understanding of spin-blockade physics in silicon quantum dots by studying 16 charge configurations of a double quantum dot coupled to a resonator. Trough magnetospectroscopy and analysis of the resonator response, we observe non-reciprocal blockade and decoherence-driven blockade lifting. Additionally, we report the largest coupling of different electron spin manifolds yet of 7.90~$\mu$eV which may enable electrical qubit control. \red{Characters: 637. Limit: 600.}

Spin qubits in gate-defined silicon quantum dots are receiving increased attention thanks to their potential for large-scale quantum computing. Readout of such spin qubits is done most accurately and scalably via Pauli spin blockade (PSB), however various mechanisms may lift PSB and complicate readout. In this work, we present an experimental observation of a new, highly prevalent PSB-lifting mechanism in a silicon double quantum dot due to incoherent tunneling between different spin manifolds. Through dispersively-detected magnetospectroscopy of the double quantum dot in 16 charge configurations, we find the mechanism to be energy-level selective and non-reciprocal for neighbouring charge configurations. Additionally, using input-output theory we report a large coupling of different electron spin manifolds of 7.90~$\mu\text{eV}$, the largest reported to date, indicating an enhanced spin-orbit coupling which may enable all-electrical qubit control. % \red{Characters: 603. Limit: 600.}

%Readout of spin qubits in gate-defined silicon quantum dots is done most accurately and scalably via Pauli spin blockade, however various mechanisms may lift the blockade and complicate readout. In this work, we expand the understanding of spin-blockade physics in silicon quantum dots by studying a 16 charge configurations of a double quantum dot coupled to a resonator. Trough magnetospectroscopy and analysis of the resonator response, we demonstrate non-reciprocal blockade and decoherence-driven blockade lifting. Additionally, we report the largest coupling of different electron spin manifolds yet of 7.90~$\mu$eV which may enable electrical qubit control. \red{Characters: 659. Limit: 600.}
\end{abstract}

\maketitle

%%%%%%%%%%%%%%%%%%%%%%   INTRODUCTION   %%%%%%%%%%%%%%%%%%%%%%
\section{Introduction}
Spin qubits in gate-defined silicon quantum dots (QDs) have emerged as a promising platform for implementing large-scale quantum computation owing to their long coherence times, compact size and ability to be operated at relatively high temperatures of 1-5~K~\cite{loss1998quantum, veldhorst2014addressable, yang2019silicon_pulse_engineering, petit2020universal, yang2020operation, camenzind2021spin}. Moreover, silicon spin qubits can be fabricated using industrial semiconductor manufacturing techniques, thereby presenting a path to high-yield large-scale device fabrication in which control electronics may also be integrated on-chip~\cite{maurand2016cmos, zwerver2021qubits, li2020flexible, xue2021cmos, pauka2019cryogenic}. With the increasing emphasis on scalability, it is becoming increasingly attractive to pursue methods of spin qubit control and readout that avoid additional on-chip components, complex device architectures, and operational constraints such as vicinity to charge reservoirs~\cite{veldhorst2017silicon, huang2017electrically, corna2018electrically, ahmed2018radio, west2019gate, crippa2019gate, pakkiam2018single}. 

Readout of silicon spin qubits is achieved by spin-to-charge conversion, a process that translates the spin degree of freedom to a selective movement of charge and that is typically performed via Pauli spin blockade (PSB) between two QDs or spin-selective tunneling to a reservoir \cite{ono2002current, elzerman2004single}. In both cases, the movement of charge can be measured using charge sensors, such as a single-electron transistor or single-electron box \cite{schoelkopf1998radio, huang2021high, ciriano2021spin}, but only PSB can be detected dispersively using resonant circuits thus alleviating the need for electrometers and reservoirs adjacent to the qubit~\cite{betz2015dispersively,zheng2019rapid}. Moreover, PSB has been shown to provide high-fidelity readout, even at low magnetic fields and elevated temperatures~\cite{zhao2019single, yang2020operation, fogarty2018integrated, harvey2018high}. 

Pauli spin blockade, however, may be lifted thus compromising readout. Lifting of PSB may occur due to direct tunneling between triplets~\cite{betz2015dispersively} or due to fast tunneling to low-energy high-spin states, such as the four-electron quintet, which may result from small valley-orbit splittings and generally dense QD energy spectra~\cite{lundberg2020spin, seedhouse2021pauli, van2018readout}. Additionally, theory predicts that fast relaxation processes happening at rates similar or faster than the readout probe frequency can lift dispersively-detected PSB~\cite{mizuta2017quantum}, and work on a charge-sensed GaAs double quantum dot (DQD) has observed spin blockade lifting due to dephasing arising from hyperfine and spin-orbit interactions~\cite{fujita2016signatures}.

In this Letter, we investigate the spin-blockade physics of a dispersively-sensed CMOS silicon DQD in 16 charge configurations and demonstrate non-reciprocal and energy-level-selective presence of PSB. We show that PSB lifting is highly prevalent and occurs due to fast decoherence, thus expanding the understanding of dispersively-detected PSB in silicon and providing a path to positive detection of different spin manifolds by selection of the readout detuning point. By analysis of the response of a resonator coupled to the DQD using input-output theory, we determine a coupling between different electron spin manifolds of 7.90~$\mu\text{eV}$, the largest reported to date. Our results indicate the presence of enhanced spin-orbit coupling (SOC) and motivate electrical spin manipulation experiments via electric-dipole spin resonance (EDSR).

%an extraordinarily large coupling of 7.90~$\mu$eV between different electron spin manifolds, thereby suggesting the presence of enhanced spin-orbit coupling (SOC) and motivating electrical spin manipulation experiments via electric-dipole spin resonance (EDSR). 

\begin{figure}
	\centering
	\includegraphics[scale=1]{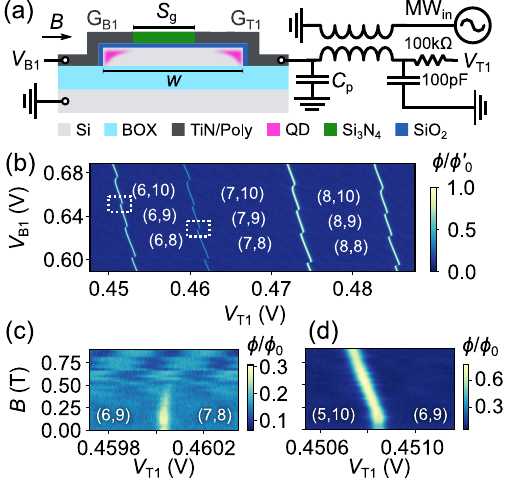}
	\caption{Non-reciprocal PSB in a pair of neighbouring ICTs. \textbf{(a)} Sketch of the device cross-section perpendicular to the direction of the nanowire illustrating the device architecture and the formation of a DQD in the upper channel corners. The $LC$ resonant circuit comprised of the DQD connected via gate $\text{G}_{\text{T1}}$ to a superconducting NbN planar spiral inductor is inductively coupled to a microstrip waveguide to enable dispersive readout. \textbf{(b)} Charge stability diagram of the DQD recorded using gate-based dispersive sensing while sweeping the electrostatic potentials of $\text{G}_{\text{T1}}$ and $\text{G}_{\text{B1}}$. The numbers in parentheses indicate the electron occupancy of the DQD and the dotted rectangles highlight the ICTs studied in panels (c) and (d) as well as in Fig.~\ref{fig2}. \textbf{(c)}, \textbf{(d)} Measurement of the relative phase shift of the (6,9)-(7,8) and (5,10)-(6,9) ICTs as a function of magnetic field $B$ showing non-reciprocal PSB.}
	\label{fig1}
\end{figure}

\section{Non-reciprocal Pauli spin blockade}
\begin{figure}
	\centering
	\includegraphics[scale=1]{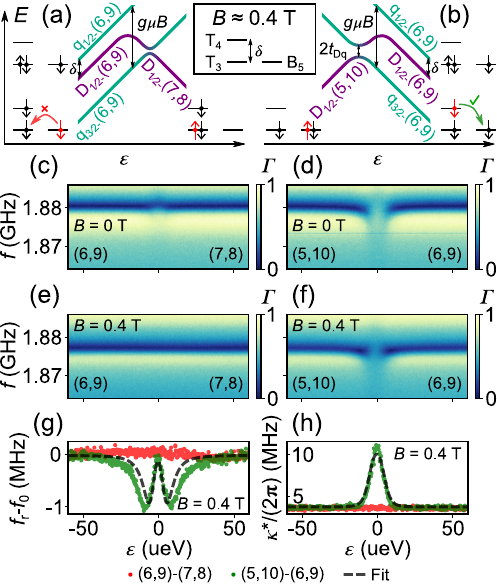}
	\caption{Level-selective PSB and decoherence. \textbf{(a), (b)} Illustrative single-particle and DQD energy levels for the (5,10), (6,9) and (7,8) charge states at $B\approx0.4$~T as a function of detuning $\varepsilon$. The red electrons and green (red) arrows indicate (the lack of) spin-flip tunneling between the doublet $\text{D}_{1/2-}$ and quadruplet $\text{q}_{3/2-}$ ground states. For simplicity, the single-particle energy levels omit the two (four) lowest-lying energy levels of $\text{QD}_\text{T1}$ $(\text{QD}_\text{B1})$. \textbf{(c)-(f)} Reflection coefficient of the resonator $\Gamma$ as a function of $f$ and $\varepsilon$ for the (6,9)-(7,8) and (5,10)-(6,9) ICTs at $B=0$~T and $B=0.4$~T. \textbf{(g),(h)} Resonance frequency and effective linewidth as a function of $\varepsilon$ for ICTs (6,9)-(7,8) [red] and (5,10)-(6,9) [green] at $B=0.4$~T determined by Lorenzian fits to the data in panels (e) and (f).}
	\label{fig2}
\end{figure}

Figure~\ref{fig1}(a) shows the split-gate nanowire field-effect transistor used in this work. To facilitate gate-based sensing~\cite{petersson2010charge}, gate $\text{G}_\text{T1}$, which we note overlaps the channel $7\pm3$~nm more than $\text{G}_\text{B1}$, is wirebonded to a superconducting NbN spiral inductor on a separate chip, thus forming a $LC$ resonant circuit with resonance frequency $f_0(B=0)=1.88$~GHz at zero magnetic field $B$. For further details about the transistor, inductor, and setup, see Appendix \ref{Sec1} and Ref. \cite{ibberson2020large}. A DQD constituted by a pair of low-symmetry QDs is formed in the upper corners of the nanowire channel when positive dc biases applied to $\text{G}_\text{B1}$ and $\text{G}_\text{T1}$ attract electrons from the source and drain reservoirs. The number of electrons accumulated in the DQD is governed by the gate voltages $V_\text{B1}$ and $V_\text{T1}$ and changes in the DQD electron occupancy are detected dispersively by probing the resonator via the transmission line $\text{MW}_\text{in}$ with a frequency $f$ close to $f_0$ and monitoring the phase shift of the reflected signal $\phi$. Non-zero phase shifts occur as a result of cyclical interdot or dot-to-reservoir tunneling events happening under the influence of the microwave probe resulting in a finite DQD-resonator coherent coupling rate $g_0$~\cite{mizuta2017quantum, ibberson2020large}. Figure~\ref{fig1}(b), which plots the phase shift $\phi$ relative to the largest dot-reservoir transition phase shift in the measured region, $\phi'_0$, as a function of $V_\text{B1}$ and $V_\text{T1}$, shows the honeycomb pattern characteristic of a DQD charge stability diagram. The electron occupancies $(N_\text{T1},N_\text{B1})$ indicated in Fig.~\ref{fig1}(b) are determined by using one QD as a charge sensor for the other QD~\cite{lundberg2020spin} (see Appendix D in Ref.~\cite{ibberson2020large} for the charge population data of the device also used in this work). While this Letter shall eventually investigate all 16 interdot charge transitions (ICTs) visible in Fig.~\ref{fig1}(b), we first consider the pair of neighbouring ICTs between the (7,8), (6,9) and (5,10) charge states highlighted by the dotted boxes in Fig.~\ref{fig1}(b). We note that the three charge states for simplicity may be considered equivalent to (3,0), (2,1) and (1,2) as seen in the single-particle diagrams of Figs.~\ref{fig2}(a) and (b).

To probe the spin physics of the DQD and in particular of the (6,9)-(7,8) and (5,10)-(6,9) ICTs, we measure $V_\text{T1}$ line traces that intersect the centre of each ICT while increasing $B$ from 0 to 0.9~T. $B$ is applied in-plane with the device and perpendicular to the nanowire, and the probe frequency $f$ is continuously adjusted to account for the changing kinetic inductance of the NbN inductor and remain close to resonance $f_0(B)$~\cite{lundberg2020spin}. The resulting dispersive magnetospectroscopy measurements shown in Figs.~\ref{fig1}(c) and \ref{fig1}(d) plot the relative phase shift $\phi/\phi_0$, where $\phi_0$ is the largest phase shift measured among the 16 ICTs in Fig.~\ref{fig1}(b). The asymmetrically vanishing signal at $B\approx0.3$~T in the (6,9)-(7,8)-magnetospectroscopy data seen in Fig.~\ref{fig1}(c) is a clear sign of dispersively detected PSB and enables us to extract an interdot lever arm of  $\alpha=0.660$ for the ICT assuming an electron $g\text{-factor}$ of 2~\cite{schroer2012radio, house2015radio, betz2015dispersively, urdampilleta2015charge, landig2019microwave, hutin2019gate, mizuta2017quantum}. Because of the shared charge state between the (6,9)-(7,8) and (5,10)-(6,9) ICTs, one might expect the (5,10)-(6,9) ICT to also show PSB~\cite{johnson2005singlet, schroer2012radio}, however, instead of vanishing from one side, the signal of this ICT, as seen in Fig.~\ref{fig1}(d) and Appendix \ref{Sec2}, persists and remains constant beyond $B=0.3$~T with a slope implying $\alpha=0.789$.

%\section{Level-selective Pauli spin blockade and decoherence} 
To facilitate the understanding of the non-reciprocal presence of PSB in Figs.~\ref{fig1}(c) and (d), we focus on the three outermost electrons in the DQD and sketch the single- and multi-particle energy levels for these electrons. In the (5,10), (6,9) and (7,8) charge states, the three outermost electrons are distributed between the $\text{T}_3$, $\text{T}_4$ and $\text{B}_5$ energy levels shown in the inset of Figs.~\ref{fig2}(a) and (b), where $\text{T}_i$ ($\text{B}_i$) refers to the $i'\text{th}$ energy level of $\text{QD}_\text{T1}$ ($\text{QD}_\text{B1}$). By populating the single-particle energy levels with the number of electrons corresponding to the charge state, we observe that the electrons in the DQD may form doublets D with one unpaired electron and a spin angular momentum $S=1/2$ or quadruplets q with three unpaired electrons and $S=3/2$. At $B=0$~T, the ground state D(6,9) is separated from the first excited state q(6,9) by $\delta=18.65~\mu\text{eV}$ (see Appendices \ref{Sec3} and \ref{Sec6} for extraction of $\delta$ and energy spectra at $B=0$~T). However, as $B$ is increased, the states Zeeman split according to $E_Z = m_\text{s}g\mu_\text{B}B$, where $m_\text{s}$ is the spin-angular-momentum projection onto the $B$-axis, $g$ is the electron $g$-factor and $\mu_\text{B}$ is the Bohr magneton, thus causing the lowest-energy quadruplet $\text{q}_{3/2-}$ to become the (6,9) ground state when $g\mu_\text{B}B>\delta$. This situation is sketched in Figs.~\ref{fig2}(a) and (b), which show the energy levels of the DQD as a function of detuning $\varepsilon$ at $B\approx0.4$~T. From these energy levels, one would expect the ground state electron transitions  $\text{q}_{3/2-}(6,9)\text{-D}_{1/2-}(7,8)$ and $\text{q}_{3/2-}(6,9)\text{-D}_{1/2-}(5,10)$, highlighted by the red electrons and arrows in Fig.~\ref{fig2}(a) and (b) respectively, to be spin-blocked due to the Pauli exclusion principle. However, recalling Fig.~\ref{fig1}, PSB is only present for the (6,9)-(7,8) transition involving levels $\text{T}_3$ and $\text{B}_5$, and not for the (5,10)-(6,9) transition, which involves levels $\text{T}_4$ and $\text{B}_5$. This hints at a level-selective process that allows spin-flip tunneling as the explanation for the signal generated at the $\text{D}_{1/2-}(5,10)\text{-q}_{3/2-}(6,9)$-intersection as a function of $B$. 

\begin{figure*}
	\centering
	\includegraphics[width=\textwidth]{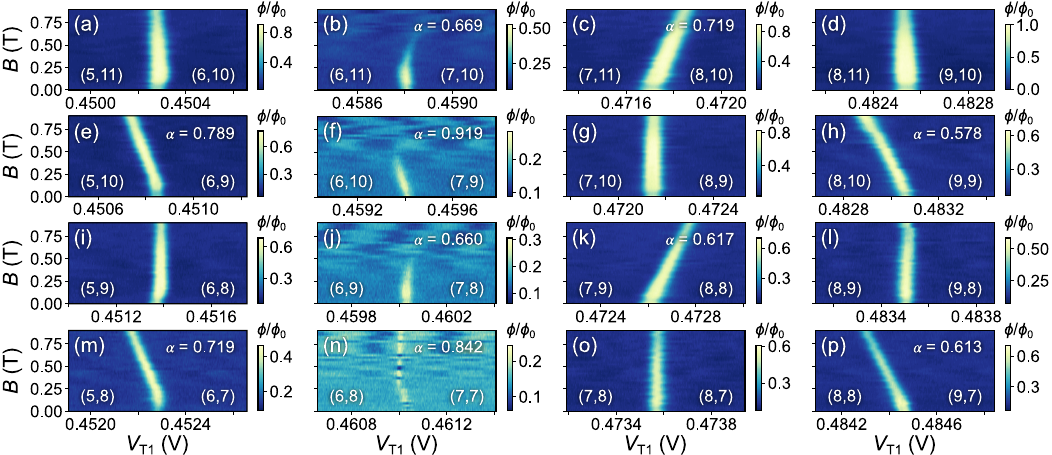}
	\caption{Magnetospectroscopy of 16 ICTs. \textbf{(a)-(p)} Measurement of the relative phase shift of the 16 ICTs visible in Fig.~\ref{fig1}(b) as a function of $B$. The numbers in parentheses indicate the electron occupancy of the DQD on either side of the ICT. Energy levels similar to those of Figs.~\ref{fig2}(a) and (b) which help explain the magnetospectroscopy are included in Appendix \ref{Sec6}.}
	\label{fig3}
\end{figure*}

\section{Pauli spin blockade lifting Mechanism}
We further investigate the two ICTs by studying the reflected spectrum of the resonator while changing detuning $\varepsilon=e\alpha (V_\text{T1}-V_\text{T1}^0)$, where $V_\text{T1}^0$ is the central position of the measured ICT, at both $B=0$~T and $B=0.4$~T [see Figs.~\ref{fig2}(c)-(f)]~\cite{petersson2012circuit}. Due to the effective coherent DQD-resonator coupling given by ${g_\text{eff}}_{ij}/(2\pi)=g_0/(2\pi) \bra{i} \hat{n} \ket{j}$ for states $i$ and $j$, where $g_0/(2\pi) = \alpha f_0 \sqrt{Z_r/2R_\text{Q}}$ in which $\alpha$ is assumed constant for a given charge configuration, $R_\text{Q}$ is the resistance quantum, $\bra{i} \hat{n} \ket{j}$ is the coupling matrix element, and $\hat{n}$ is the charge number operator, we observe changes in the resonance frequency $f_r$ and effective linewidth $\kappa^{*}$ of the resonator according to~\cite{childress2004mesoscopic, koch2007charge, ibberson2020large}
\begin{equation}
	f_r = f_0 - \frac{1}{2\pi}\frac{{g_\text{eff}}_{ij}^2\Delta_{ij}}{\Delta_{ij}^2+\gamma_{ij}^2/4} 
	\label{eq1}
\end{equation}
\begin{equation}
	\kappa^{*}/(2\pi) = \kappa/(2\pi) + \frac{1}{2\pi} \frac{{g_\text{eff}}_{ij}^2\gamma_{ij}}{\Delta_{ij}^2+\gamma_{ij}^2/4}
	\label{eq2} 
\end{equation}
where $t_{ij}$ is the DQD tunnel coupling, $\Omega_{ij}=\sqrt{\varepsilon^2+4t_{ij}^2}$ is the energy difference between participating states, $\Delta_{ij} = \Omega_{ij}/\hbar-2\pi f_0$, $\kappa$ is the bare resonator linewidth, and $\gamma_{ij}$ is the DQD decoherence rate. By applying these equations to the measurements in Figs.~\ref{fig2}(c)-(f), we can deduce important information about the DQD.

Inspection of the resonator response at $B=0$~T in Fig.~\ref{fig2}(c) reveals an upwards shift of $f_r$ around $\varepsilon=0$ for the (6,9)-(7,8) transition thus indicating ${g_\text{eff}}_\text{DD}\neq~0, \Delta_\text{DD} < 0$ and hence the tunnel coupling $2t_\text{DD} < h f_0 = 7.8~\mu\text{eV}$ for states $\text{D}(6,9)\text{ and D}(7,8)$. Despite being in the resonant regime at $\varepsilon$ where $\Omega_\text{DD}=hf_0$, we do not observe characteristic vacuum Rabi-mode-splitting due to the large $\gamma_\text{DD}$ of this charge transition indicated by the increased $\kappa^{*}$ around $\varepsilon=0$ (see Appendix \ref{Sec5}). By similar inspection of Fig.~\ref{fig2}(d), we find a downwards shift of $f_r$ around $\varepsilon=0$ indicating that ${g_\text{eff}}_{\text{DD}'}\neq~0, \Delta_{\text{DD}'}>0$ and hence that the tunnel coupling $2t_{\text{DD}'} > h f_0 = 7.8~\mu\text{eV}$ for states $\text{D}(5,10)\text{ and D}(6,9)$. To better understand the origin of non-reciprocal PSB, we repeat the resonator response measurements at $B=0.4$~T as shown in Figs.~\ref{fig2}(e) and (f) and furthermore fit each $\varepsilon$ trace of the response to a Lorenzian with centre frequency $f_r$ and linewidth $\kappa^{*}/(2\pi)$ (see Appendix \ref{Sec4} for description of fitting procedure). As can be seen from the red (6,9)-(7,8) data in Figs.~\ref{fig2}(g) and (h), there are no changes to $f_r$ or $\kappa^{*}$ around $\varepsilon=0$, thus indicating that the coupling matrix element $\langle\text{q}_{3/2-}(6,9) | \hat{n} | \text{D}_{1/2-}(7,8)\rangle=0$. This conclusion agrees with the lack of phase response at $B\geq0.4$~T due to PSB seen in Fig.~\ref{fig1}(c). However, for the (5,10)-(6,9) transition represented by the green data in Figs.~\ref{fig2}(g) and (h), the shift in $f_r$ and increase in $\kappa^{*}$ around $\varepsilon=0$ indicates ${g_\text{eff}}_\text{Dq}\neq0$ for states $\text{D}_{1/2-}(5,10)\text{ and q}_{3/2-}(6,9)$. The observation that $f_r$ first decreases symmetrically with decreasing $|\varepsilon|$ indicates $\Delta_\text{Dq} > 0$ whereas the subsequent increase to $f_r\approx f_0$ at $\varepsilon=0$ indicates $\Delta_\text{Dq} \rightarrow 0$ when $\varepsilon \rightarrow 0$. This implies a $\text{D}_{1/2-}(5,10)\text{-q}_{3/2-}(6,9)$ tunnel coupling of $2t_\text{Dq}\approx hf_0 = 7.8~\mu\text{eV}$, which is confirmed by fitting $f_r-f_0$ and $\kappa^{*}$ simultaneously with shared parameters to Eqs. \eqref{eq1} and \eqref{eq2} from which we obtain $2t_\text{Dq}=7.90\pm0.01~\mu\text{eV}$, $g_0/(2\pi)=51.0\pm0.2~\text{MHz}$, and $\gamma_\text{Dq}/(2\pi)=1.65\pm0.01~\text{GHz}$. For electrons in silicon, $7.90~\mu\text{eV}$ is a remarkably large tunnel coupling between states with different $S$, more than twice the value theoretically estimated in Corna \textit{et al}~\cite{corna2018electrically}. 

The spin-flip (5,10)-(6,9) charge transition at $B>0.3~\text{T}$, confirmed by the non-zero coupling matrix element, is identified as an incoherent tunneling process due to the large increase in the resonator linewidth around $\varepsilon=0$. The large $\gamma_\text{Dq}/(2\pi)=1.65\pm0.01~\text{GHz}$ indicates that the system decoheres on the timescale of the resonator and hence that the measurable signal at the $\text{D}_{1/2-}\text{-q}_{3/2-}$ anti-crossing results from the net charge transfer of multiple incoherent passages. We therefore discard coherent tunneling by adiabatic passage as the origin of the signal, despite the large $t_\text{Dq}$. Incoherent tunneling may be the consequence of fast state relaxation or pure dephasing. We can further narrow down the origin of the signal by discarding ordinary intradot valley-hotspot-mediated relaxation~\cite{yang2013spin} since this process is highly magnetic field-dependent and hence incompatible with the relatively constant signal seen in Fig.~\ref{fig1}(d) and in Appendix \ref{Sec2}. Whether decoherence is dominated by interdot relaxation or pure dephasing remains to be determined, but given that decoherence is generally dephasing-limited in silicon spin systems we speculate the latter to be the dominant mechanism. The large value of $t_\text{Dq}$ indicates the presence of significant SOC which may open dephasing channels similar to those observed for spin-preserving transitions in corner QDs~\cite{dupont2013coherent,gonzalez2016gate,chatterjee2018silicon} that are compatible with our measured value of $\gamma_\text{Dq}$. Additionally, spin, valley, and orbital degrees of freedom may all be mixed on a wide energy range by spin-orbit and valley-orbit couplings, which together with Coulomb interactions provide efficient paths for decoherence. In particular, it has been shown in \cite{abadillo2021wigner} that anisotropic QDs such as the ones characterised in this work are prone to Wigner-like localization: The electrons split apart in the charged dots due to Coulomb repulsion, which results in a significant compression of the energy spectrum~\cite{lundberg2020spin} and mixing of the different degrees of freedom in the presence of spin-orbit and valley-orbit coupling mechanisms~\cite{ercan2021strong, ercan2021charge, corrigan2021coherent}. Although the observed $\gamma_\text{Dq}$ at the anti-crossing is large, one may expect a reduction away from the anti-crossing where the energy difference of intradot transitions is $\varepsilon$-independent, thus encouraging coherent EDSR experiments for electron spins in silicon corner dots. Additionally, we note that the spin-blockade-lifting mechanism presented here does not render spin readout impossible but does lead to a reduced and magnetic-field-dependent readout window. In fact, the dispersive signals generated at the anti-crossing points of the different spin manifolds enable positive detection of both the singlet (doublet) at $\varepsilon=0$ and the triplet $\text{T}_\pm$ (quadruplet $\text{q}_{3/2\pm}$) at $\varepsilon=\pm g\mu_\text{B}B$ through selection of the detuning value at which readout is performed. 

\section{Prevalence of Pauli spin blockade}
To better understand the prevalence of PSB or lack thereof, we expand the magnetospectroscopy measurements to all 16 ICTs visible in Fig.~\ref{fig1}(b). The resulting measurements are shown in Fig.~\ref{fig3} in which panels (e) and (j) represent the measurements studied in Fig.~\ref{fig1}. The measurements fall into four categories, which are covered in detail in Appendix \ref{Sec6} and summarised here: Most abundant for this DQD is the lack of PSB due to fast spin-flip transitions as seen in Figs.~\ref{fig3}(c), (e), (f), (h), (k), (m), (n), and (p), among which (f) and (n) are unusual due to the small splitting $\delta$ as explained in Appendix \ref{Sec6}. Regular PSB on the other hand is only seen in panels (b) and (j), and partial PSB in which large tunnel couplings relative to $\delta$ obscure the blockaded region (see Appendices \ref{Sec6} and \ref{Sec7}) found in panels (a) and (i). Finally, Figs.~\ref{fig3}(d), (g), (l), and (o) demonstrate cases without PSB in which the spin ground state remains the same for all measured $B$ consistent with previously reported odd-parity transitions~\cite{schroer2012radio}. Besides the prevalence, we find that the measurements have a periodicity of two in the $\text{QD}_\text{B1}$ occupancy, indicating broken valley degeneracy and large level separation in $\text{QD}_\text{B1}$. This agrees with the addition energy that can be inferred from the charge stability diagram and the smaller QD size expected from the lesser channel overlap of $\text{G}_\text{B1}$. From the sloped magnetospectroscopy measurements in Fig.~\ref{fig3}, we extract $\alpha$ and find a consistently large average $\alpha_\text{ave}=0.7\pm0.1$ across the ten ICTs involving different spin manifolds.

\begin{figure}
	\centering
	\includegraphics[scale=1]{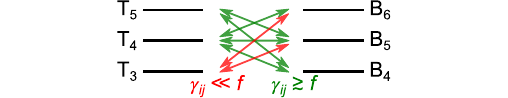}
	\caption{Energy level combinations with fast and slow decoherence. \textbf{(a)} Single-particle energy levels for $\text{QD}_\text{T1}$ and $\text{QD}_\text{B1}$ connected by arrows which show whether the indicated ICT has small or large $\gamma_{ij}$ relative to $f$. The two (three) lowest energy levels of $\text{QD}_\text{T1}$ ($\text{QD}_\text{B1}$)  are omitted for simplicity. }
	\label{fig4}
\end{figure}

By analysis and simulation of each measurement presented in Fig.~\ref{fig3}, we construct energy levels similar to those in Figs.~\ref{fig2}(a) and (b) (see Appendices \ref{Sec6} and \ref{Sec7}), which enable us to understand which combinations of QD energy levels produce fast spin-flipping. The resulting analysis is summarised in Fig.~\ref{fig4}, which shows that all transitions except those that involve level $\text{T}_3$ have $\gamma_{ij} \gtrsim f$. The reason for the level-selectivity is unknown, but may be linked to the orbital and valley quantum numbers of the QD states involved. 

% std {669, 719, 789, 919, 578, 660, 617, 719, 824, 613}

%%%%%%%%%%%%%%%%%%%%%%   CONCLUSIONS AND OUTLOOK   %%%%%%%%%%%%%%%%%%%%%%
\section{Conclusions and outlook}
In summary, we have studied the spin-blockade physics of a silicon electron DQD coupled to a resonator probed in reflectometry and through dispersive magnetospectroscopy measurements of 16 ICTs found the presence of PSB to be non-reciprocal and energy-level selective. By analysing the resonator response as a function of DQD detuning and probe frequency using input-output theory, we reported the largest tunnel coupling between different electron spin manifolds yet $2t_\text{Dq}=7.90~\mu\text{eV}$ and determined that the lifting of PSB is due to fast decoherence $\gamma_\text{Dq}$ at the manifold anti-crossing. This finding provides a way of positively detecting both orientations of a spin, e.g. singlets and triplets, through selection of appropriate readout detuning values and may thus enhance readout fidelities. The large $t_\text{Dq}$, possibly due to enhanced spin-orbit coupling of silicon corner QDs, encourages attempts to perform EDSR away from the anti-crossings followed by readout at the chosen anti-crossing. Overall, our results build a new understanding of PSB physics in silicon important to spin qubit readout and motivate pursuit of all-electrical control of electron spin qubits.

%%%%%%%%%%%%%%%%%%%%%%   ACKNOWLEDGEMENTS   %%%%%%%%%%%%%%%%%%%%%%
\begin{acknowledgements}
We thank M. Benito and C. Lainé for valuable discussions. This research has received funding from the European Union's Horizon 2020 Research and Innovation Programme under grant agreements No 688539 and 951852. T.L. acknowledges support from the EPSRC Cambridge NanoDTC, EP/L015978/1. D.J.I. is supported by the Bristol Quantum Engineering Centre for Doctoral Training, EPSRC Grant No. EP/L015730/1. M.F.G.Z. acknowledges support from Industrial Strategy Challenge Fund, UKRI. C.L. and J.W.A.R.  acknowledge the EPSRC through the Core-to-Core International Network “Oxide Superspin” (EP/P026311/1) and the “Superconducting Spintronics” Programme Grant (EP/N017242/1).
\end{acknowledgements}

\appendix
\section{Description of device and measurement setup} \label{Sec1}
Figure~\ref{fig1}(a) depicts the formation of a DQD in a split-gate nanowire field-effect transistor as well as the resonant circuit used to perform dispersive gate-sensing of the DQD. The transistor, which is fabricated on a 300-mm fully-depleted silicon-on-insulator wafer with a buried oxide (BOX) thickness of 145~nm, consists of a $h=7$~nm thick, $w=70$~nm wide silicon nanowire channel, a 6~nm SiO$_2$ gate oxide and two TiN/polysilicon gates, G$_\text{B1}$ and G$_\text{T1}$, with 60~nm gate lengths and a $S_\text{g}=40$~nm split-gate separation. We note that G$_\text{T1}$ overlaps the channel slightly more than G$_\text{B1}$ due to a $7\pm3$~nm misalignment during fabrication. Spacers made of Si$_3$N$_4$ are used to extend the region of intrinsic silicon beneath the gates by 34~nm on either side, thus creating tunnel barriers to the heavily \textit{n}-type-doped source and drain that are held at 0~V. To facilitate gate-based sensing, G$_\text{T1}$ is wirebonded to a superconducting NbN spiral inductor on a separate chip, thus forming a $LC$ resonant circuit comprised of the capacitance $C_\text{d}$ from G$_\text{T1}$ to ground, parasitic capacitance $C_\text{p}$, and magnetic-field-dependent inductance $L(B)$ of the spiral. At $B=0$~T, the resonator has a resonance frequency of $f_0(B=0)=1.88$~GHz, however when the magnetic field $B$ is increased, the resonance frequency $f_0(B)=1/[2\pi \sqrt{L(B)(C_\text{d}+C_\text{p})}]$ decreases due to the increasing kinetic inductance of the superconducting spiral inductor~\cite{lundberg2020spin}. The inductor, which is fabricated by optical lithography of an 80-nm-thick sputter-deposited NbN film on a sapphire substrate, is situated adjacent to a 50~$\Omega$ microstrip waveguide and designed to achieve critical coupling between the resonator and input line. This design results in a large resonator characteristic impedance $Z_r=560~\Omega$ which, together with the large lever arm $\alpha=C_\text{T1,T1}/C_{\Sigma\text{T1}} - C_\text{T1,B1}/C_{\Sigma\text{B1}}\approx0.7$ of the wrap-around gates (where $C_{i,j}$ is the capacitance between gate $i$ and QD $j$ and $C_{\Sigma i}$ is the total capacitance of QD $i$), enables a large coherent coupling rate $g_0$ and thus a large signal-to-noise ratio~\cite{ibberson2020large}. We probe the resonator-DQD system in reflectometry via the microwave transmission line labelled MW$_\text{in}$ in Fig.~\ref{fig1}.

\section{Maximum phase shift of (5,10)-(6,9) ICT vs magnetic field}  \label{Sec2}
\begin{figure}[h]
	\centering
	\includegraphics[width=0.5\textwidth]{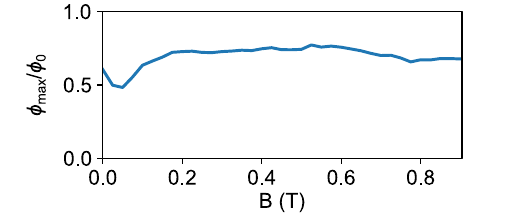}
	\caption{Maximum relative phase shift for each $B$-field linetrace of the magnetospectroscopy measurement presented in Figs.~\ref{fig1}(d) and \ref{fig3}(e). The decrease in relative phase shift around $B=45$~mT is due to a decrease in the resonator Q-factor at this magnetic field.}
	\label{sfig1}
\end{figure}

To further characterise the lack of Pauli spin blockade (PSB) observed for the (5,10)-(6,9) ICT in the magnetospectroscopy measurement of Figs.~\ref{fig1}(d) and \ref{fig3}(e), we extract the maximum relative phase shift $\phi_\text{max}/\phi_0$, where $\phi_0$ is the largest phase shift measured among the 16 ICTs in Fig.~\ref{fig1}(b), for each $B$-field linetrace as seen in Fig.~\ref{sfig1}. With the exception of a small decrease around $B=45$~mT, Fig.~\ref{sfig1} shows that $\phi_\text{max}/\phi_0$ remains relatively constant from $B=0$ to $B=0.9$. As explained in the main text, the constant phase shift rules out magnetic-field-dependent processes such as standard intradot valley-mediated-relaxation~\cite{yang2013spin} as the explanation for the lack of PSB. We note that the decreased phase shift around $B=45$~mT is also observed in magnetospectroscopy measurements of all the other ICTs in Fig.~\ref{fig3}. We attribute the decrease in phase shift to a reduction of the resonator Q-factor around $45$~mT. This deterioration of the Q-factor is also observed in measurements of the reflected spectrum of the resonator as a function of magnetic field during which the device is grounded.

\begin{figure}
	\centering
	\includegraphics[width=0.5\textwidth]{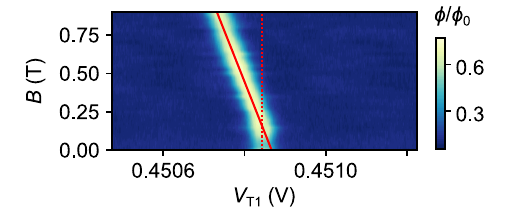}
	\caption{Measurement of the relative phase shift of the (5,10)-(6,9) ICT as a function of magnetic field $B$. The dotted red line marks the central position of the phase shift signal at $B=0$~T, whereas the solid red line fits the sloped region for $B>0.2$~T and is extrapolated to $B=0$~T.}
	\label{sfig2}
\end{figure}

\section{Extracting T$_3$-T$_4$ energy separation} \label{Sec3}
As can be seen from Fig.~\ref{fig2}, the D(6,9)-q(6,9) energy separation $\delta$ is given by the splitting of levels T$_3$ and T$_4$. The magnitude of $\delta$ may be extracted from the magnetospectroscopy measurement presented in Fig.~\ref{sfig2} [same data as shown in Figs.~\ref{fig1}(d) and \ref{fig3}(e)] by extrapolating the sloped signal of the D(5,10)-q(6,9) anti-crossing to $B=0$~T and determining the voltage separation of the extrapolated signal from the doublet anti-crossing signal at $B=0$~T. To extrapolate the signal of the D(5,10)-q(6,9) anti-crossing, we extract the centre of the phase shift peak for each line trace above $B>0.2$~T and linearly fit these points to obtain the solid red line in Fig.~\ref{sfig2}. From the intersection of the linear fit with $B=0$~T and the central position of the phase shift signal at $B=0$~T marked with a dotted red line in Fig.~\ref{sfig2}, we find a voltage separation of $\Delta V = 23.63~\mu\text{V}$, which entails $\delta = \alpha \Delta V = 0.789\cdot23.63~\mu\text{V}=18.65~\mu\text{eV}$.

\section{Resonator response for ICTs (6,9)-(7,8) and (5,10)-(6,9) at \textit{B}=0~T} \label{Sec5}
\begin{figure}[h]
	\centering
	\includegraphics[width=0.5\textwidth]{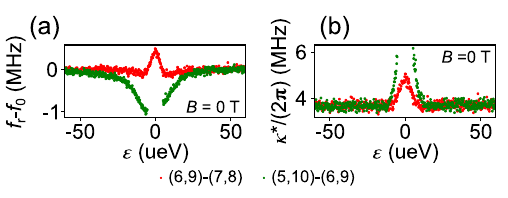}
	\caption{\textbf{(a), (b)} Resonance frequency $f_r$ and effective linewidth $\kappa^*$ as a function of $\varepsilon$ for ICTs (6,9)-(7,8) [red] and (5,10)-(6,9) [green] at $B=0$~T determined by Lorenzian fits to the data in Figs.~\ref{fig2}(c) and (d).}
	\label{sfig3}
\end{figure}

To emphasise the features of the resonator response at $B=0$~T shown in Figs.~\ref{fig2}(c) and (d), we perform the same analysis and Lorentzian fitting as for Figs.~\ref{fig2}(e) and (f) and as is described in Appendix~\ref{Sec4}. The $f_r$ and $\kappa^*/(2\pi)$ values obtained for the fits are plotted in Fig.~\ref{sfig3}. From Fig.~\ref{sfig3}(a) it is immediately clear that the resonance shifts up [down] for ICT (6,9)-(7,8) [(5,10)-(6,9)], thus providing further evidence for the observations made in the main text that $t_\text{DD}<hf_0$ and $t_{\text{DD}'}>hf_0$. From Fig.~\ref{sfig3}(b), we note a sizeable increase in $\kappa^*$ around $\varepsilon=0$ indicating a large decoherence rates $\gamma_{\text{DD}}$ and $\gamma_{\text{DD}'}$ for the (6,9)-(7,8) and (5,10)-(6,9) doublet transitions. This explains the absence of vacuum Rabi-mode-splitting, which may otherwise have been expected for the resonant regime of ICT (6,9)-(7,8) at $B=0$~T. The large $\gamma_{\text{DD}'}$ also complicates fitting the resonator response of ICT (5,10)-(6,9) around $\varepsilon=0$ to Eq. \eqref{S11} and is therefore the reason for the lacking data points around $\varepsilon=0$ in Fig.~\ref{sfig3}. We note that $t_\text{DD}$ is smaller than $t_\text{Dq}$ indicating that in this device coupling between states with the same spin angular momentum $S$ can be smaller than between states with different $S$. 

\section{Description of fitting procedures} \label{Sec4}
For the analysis of the resonator response as a function of detuning $\varepsilon$, we use the steady-state power reflection coefficient developed from the Heisenberg-Langevin equations of motion in its complex Lorentzian form~\cite{ibberson2020large}
%\begin{equation}
%	\left\lvert S_{11} \right\rvert = \left\lvert 1-\frac{i\kappa_\text{ext}/(2\pi)}{f - f_0+\frac{1}{2\pi}\frac{g_\text{eff}^2\Delta}{\Delta^2+\gamma^2/4}+\frac{i}{4\pi} \left(\kappa+\frac{g_\text{eff}^2\gamma}{\Delta^2+\gamma^2/4} \right)} \right\rvert
%\end{equation}
\begin{equation}
	\left\lvert S_{11} \right\rvert = \left\lvert 1-\frac{i\kappa_\text{ext}/(2\pi)}{f - f_r+\frac{i}{2} \kappa^*/(2\pi)} \right\rvert^2
	\label{S11}
\end{equation}
where $\kappa_\text{ext}/(2\pi)=1.76$~MHz is the external photon decay rate, and where $f_r$ as well as $\kappa^*/(2\pi)$ are defined in Eqs. (1) and (2) in the main text. To obtain the values for $f_r$ and $\kappa^*/(2\pi)$ plotted in Figs.~\ref{fig2}(g) and (h) as well as Figs.~\ref{sfig3}(a) and (b) below, we fit each $\varepsilon$ line trace of the data presented in Figs.~\ref{fig2}(c)-(f) to Eq.~\eqref{S11}.

We use the $f_r$ and $\kappa^*/(2\pi)$ data at $B=0.4$~T plotted in Figs.~\ref{fig2}(g) and (h) to extract the coherent coupling rate $g_0$, tunnel coupling $t_\text{Dq}$, and decoherence rate $\gamma_\text{Dq}$. This is done by simultaneously fitting $f_r$ and $\kappa^*/(2\pi)$ to Eqs. (1) and (2) in the main text with shared fitting parameters $g_0$, $t_\text{Dq}$, and $\gamma_\text{Dq}$ and using orthogonal distance regression (ODR) that factors in the errors on both axis for each dataset. The errors given for $g_0$, $t_\text{Dq}$, and $\gamma_\text{Dq}$ in the main text are obtained from the covariance matrix of the ODR fit and represent one standard deviation.

\section{Energy-level diagrams for all 16 ICTs at \textit{B}=0~T and \textit{B}=0.4~T} \label{Sec6}
Figures~\ref{sfig4} and \ref{sfig5} sketch the single-particle and DQD energy levels as a function of detuning $\varepsilon$ at $B=0$~T and $B=0.4$~T, respectively, for all 16 ICTs visible in the charge stability diagram of Fig.~\ref{fig1}(b). The single-particle energy levels shown in each panel of Figs.~\ref{sfig4} and \ref{sfig5} represent the six energy levels shown in Fig.~\ref{fig4}, which means that the two (three) lowest and fully-occupied energy levels of QD$_\text{T1}$ (QD$_\text{B1}$) have been omitted for simplicity. The electrons highlighted in red in the single-particle energy diagrams indicate which electrons move QD when crossing an ICT, whereas the presence of a green (red) arrow indicates the presence (lack) of spin-flip tunneling. The colour of the DQD energy levels indicate the multiplicity of the multi-particle spin state it represents, and for simplicity, couplings between manifolds of different spin angular momentum are not included in the sketch. To strengthen the link to the main text, the panels of Figs.~\ref{sfig4} and \ref{sfig5} are organised such that they match the location of the ICTs in the charge stability diagram as well as the panel labels of Fig.~\ref{fig3}. This means that one electron is added to QD$_\text{T1}$ (QD$_\text{B1}$) when moving one panel to the right (up). One may note from the first and second column of panels in Figs.~\ref{sfig4} and \ref{sfig5}, that the small splitting $\delta$ is a defining feature which, for example, explains the presence of the low-energy quadruplet states. In the following, we split the energy diagrams into groups with similar features and use them as a basis for explaining the corresponding magnetospectroscopy panels of Fig.~\ref{fig3}:

\begin{figure*}[h]
	\centering
	\includegraphics[width=0.96\textwidth]{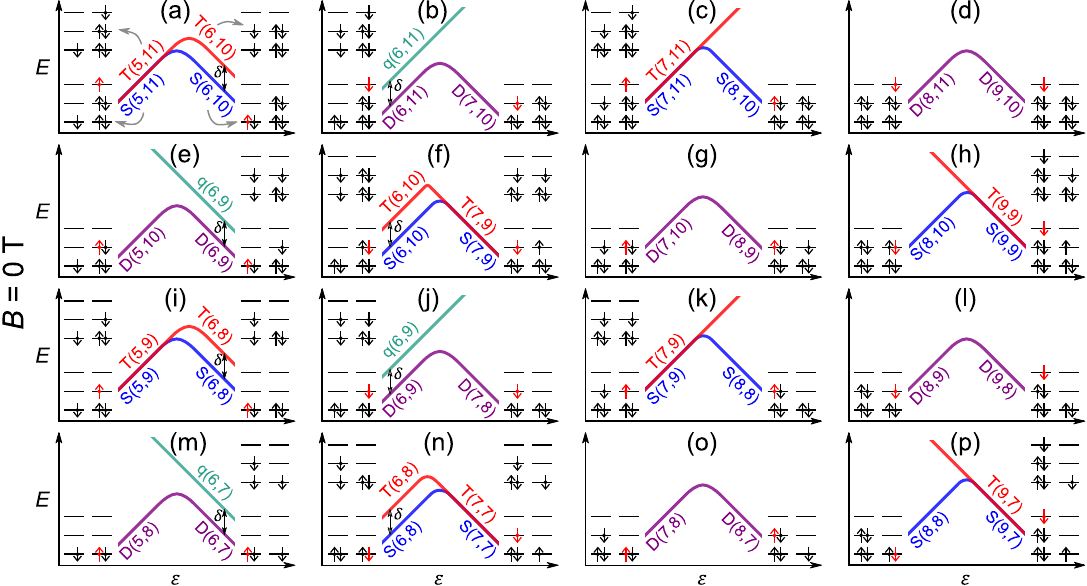}
	\caption{Illustrative single-particle and DQD energy levels as a function of detuning $\varepsilon$ at $B=0$~T for the charge states involved in the 16 ICTs present in the charge stability diagram of Fig.~\ref{fig1}(b). For simplicity, the single particle energy levels omit the two (three) lowest-lying energy levels of QD$_\text{T1}$ (QD$_\text{B1}$), such that the six energy levels shown are T$_3$, T$_4$, T$_5$, B$_4$, B$_5$, B$_6$ starting from the bottom left. The red electron indicates the electron that moves QD as a function of changes in $\varepsilon$.}
	\label{sfig4}
\end{figure*}

\begin{figure*}[h]
	\centering
	\includegraphics[width=0.96\textwidth]{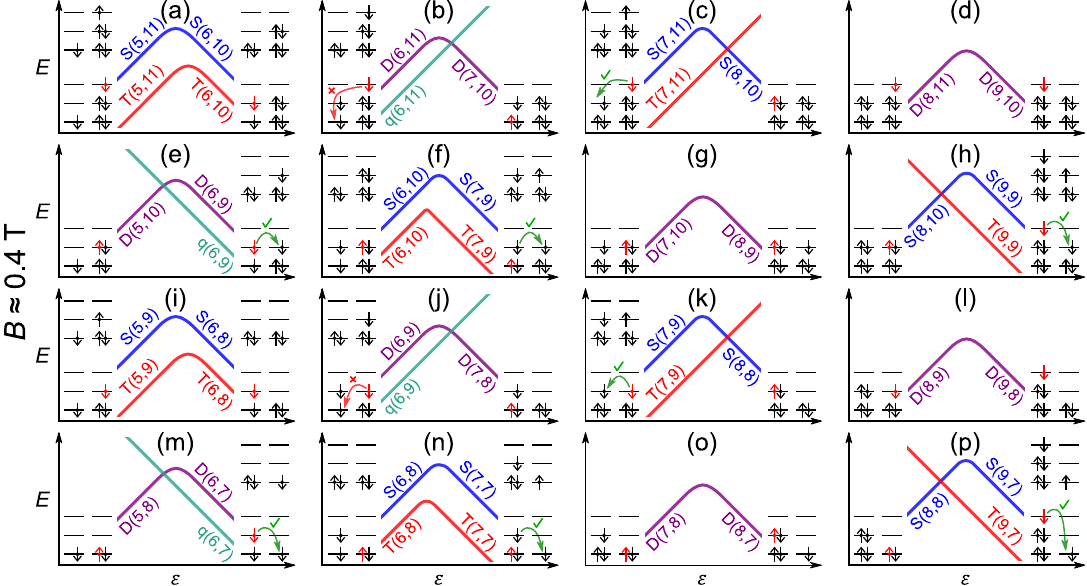}
	\caption{Illustrative single-particle and DQD energy levels as a function of detuning $\varepsilon$ at $B=0.4$~T for the charge states involved in the 16 ICTs present in the charge stability diagram of Fig.~\ref{fig1}(b). For simplicity, the single particle energy levels omit the two (three) lowest-lying energy levels of QD$_\text{T1}$ (QD$_\text{B1}$), such that the six energy levels shown are T$_3$, T$_4$, T$_5$, B$_4$, B$_5$, B$_6$ starting from the bottom left. The red electron indicates the electron that moves QD as a function of changes in detuning, and the green (red) arrows indicate (the lack of) spin-flip tunneling. Note that the green arrows in panels (f) and (n) indicate spin-flip tunneling in the singlet-triplet manifold which happens only when $B\lesssim\delta/(g\mu_\text{B})=0.16$~T.}
	\label{sfig5}
\end{figure*}

\subparagraph{Panels (c), (e), (f), (h), (k), (m), (n), and (p)}
As highlighted in the main text and as seen in the case of Fig.~\ref{fig3}(c), (e), (f), (h), (k), (m), (n), and (p), the lack of PSB is most abundant among the 16 ICTs. For panels (c), (h), (k), and (p), the phase shift signal of the magnetospectroscopy slopes to one side already from $B>0$~T. This indicates that the signal at $B=0$ arises from the anti-crossing of singlets [see Fig.~\ref{sfig4}(c), (h), (k), and (f)], but that the singlet tunnel coupling is small such that the signal at $B>0$~T arises from incoherent tunneling between the singlet and triplet states as indicated by the green arrows in Fig.~\ref{sfig5}(c), (h), (k), and (f).

Panels (e) and (m) are similar to panels (c), (h), (k), and (p), but involve doublet and quadruplet spin states instead of singlets and triplets. Whereas the singlets and triplets were degenerate in one charge configuration at $B=0$~T, the minimal energy difference between the doublet and quadruplet is given by $\delta$. The signal at $B=0$~T from the doublet anti-crossing [see Fig.~\ref{sfig4}(e) and (m)] therefore persists for longer, and only when $B\gtrsim\delta/(g\mu_\text{B})=0.16$~T do the quadruplets start to intersect the doublets, resulting in the characteristic sloped signal arising from incoherent tunneling.

Panels (f) and (n) start off similar to panels (c), (h), (k), and (p) with the explanation for what happens in the low-field region of the magnetospectroscopy being the same. However, as seen in Fig.~\ref{sfig4}(f) and (n), the maximal separation of the triplets from the singlets is $\delta$, which means that the triplets become the ground state when $B\gtrsim0.16$~T. The straight high-field signal in Fig.~\ref{fig3}(n) may therefore attributed to the anti-crossing triplet ground state [see Fig.~\ref{sfig5}(n)], whereas the lack of any high-field signal in Fig.~\ref{fig3}(f) may be explained by the tunnel coupling of the anti-crossing triplet states being so small that the resonator probe drives diabatic Landau-Zener transitions across the anti-crossing~\cite{nielsen2013six}. 

\subparagraph{Panels (b) and (j)}
The only two cases of clear PSB are found in panels (b) and (j), which involve doublet and quadruplet states, and whose magnetospectroscopy matches previous observations of dispersively-detected PSB~\cite{schroer2012radio, betz2015dispersively, urdampilleta2015charge, landig2019microwave, hutin2019gate}. Similar to panels (e) and (m), the low-field signal arises from the doublet anti-crossing [see Figs.~\ref{sfig4}(b) and (j)], and only when $B\gtrsim0.16$~T does PSB begin to manifest because spin-flip tunneling is prohibited for energy-level transitions T$_3$-B$_5$ and T$_3$-B$_6$ as shown with the red arrows in Figs.~\ref{sfig5}(b) and (j).

\subparagraph{Panels (a) and (i)}
Panels (a) and (i) bear similarities with panels (f) and (n) because the triplet anti-crossing also here becomes the ground state already when $B\gtrsim0.16$~T, giving rise to the high-field signal [see Figs.~\ref{sfig5}(a) and (i)]. However, because the tunnel coupling of both the singlet and triplet manifolds is large compared to $\delta$ for panels (a) and (i), we do not observe a region of no-signal PSB that would otherwise have expected from the energy-level transitions T$_3$-B$_5$ and T$_3$-B$_6$. We therefore classify these two panels as cases of partial PSB. 

\subparagraph{Panels (d), (g), (l), and (o)}
Finally, panels (d), (g), (l), and (o) represent simple cases of odd-parity transitions~\cite{schroer2012radio} where the anti-crossing of just one spin manifold, the doublet, is the ground state for the entire range of magnetic fields studied. The lack of any higher spin manifolds indicates that the energy-level separation between levels T$_4$ and T$_5$ as well as the separation between levels in QD$_\text{B1}$ greater than $g \mu_\text{B} \cdot 0.9~\text{T}\approx100~\mu \text{eV}$, i.e. substantially larger than $\delta$.

% Changes needed: 1) Make level separation the same in figs a-n. 2) Small tunnel coupling triplets in f. 3) small tunnel coupling in c, h, k, p. 4) Triplet is ground state at 0.4T in f and n. 5) Triplets in c, h, k, p should be lower than quadruplets. 6) a and i bigger tunnel coupling than f and n, but Otherwise similar.
	
\section{Magnetospectroscopy simulations for all 16 ICTs} \label{Sec7}
To support the explanations of the observed magnetospectroscopy provided in Appendix~\ref{Sec6}, we qualitatively simulate the resonator response as a function of magnetic field for the energy spectra in Figs.~\ref{sfig4} and \ref{sfig5}. We use the semi-classical approximation~\cite{mizuta2017quantum} where the effect of the quantum system on a classical resonator is expressed in terms of the parametric capacitance
\begin{equation}
	C_\text{pm}=(e\alpha)^2\dfrac{\partial \langle n_{2} \rangle}{\partial\varepsilon}
	\label{Cpm}
\end{equation}
\noindent where $e$ is the electron charge, $\alpha$ is the interdot lever arm, $\langle n_{2} \rangle$ is the occupation probability of the QD connected to the resonator. We further consider the small signal regime where the phase response of the resonator is directly proportional to the change in parametric capacitance, $\phi\propto\Delta C_\text{pm}$. The QD occupation probability can be expanded in terms of the polarization and occupation probabilities of each individual eigenstate, $\langle n_{2} \rangle_i$ and $P_i$ respectively 
\begin{equation}
	\langle n_{2} \rangle=\sum_i \langle n_{2} \rangle_iP_i.
	\label{polarization}
\end{equation}
In this case, the parametric capacitance can be expanded in terms of its two principal constituents, the quantum capacitance and tunneling capacitance~\cite{mizuta2017quantum, esterli2019small}:
\begin{equation}
	C_\text{pm}=(e\alpha)^2\sum_i \underbrace{\dfrac{\partial \langle n_{2} \rangle_i}{\partial\varepsilon}P_i \rule[-15pt]{0pt}{1.7pt}}_{\mbox{\footnotesize quantum}}+ \underbrace{\langle n_{2} \rangle_i \dfrac{\partial P_i}{\partial\varepsilon} \rule[-15pt]{0pt}{1.7pt}}_{\mbox{\footnotesize tunnelling}}.
	\label{Cpm2}
\end{equation}
Finally, to perform the simulations we consider the general expression of the polarizations:
\begin{equation}
	\langle n_{2} \rangle_i=\frac{1}{2}\left(1+2\frac{\partial E_i}{\partial\varepsilon}\right).
	\label{polarization2}
\end{equation}
To produce the simulations, the parameters that are needed are the couplings between eigenenergies in the same spin branch $t_i$, the energy splittings at large absolute detuning $\delta_i$ and the temperature $T$. In each panel, we either use the slow relaxation regime, where only quantum capacitance manifests, or the fast relaxation regime, where both quantum and tunneling capacitance contribute to the total parametric capacitance~\cite{mizuta2017quantum}. The latter implies that we are simulating fast decoherence via relaxation and hence we use the thermal probabilities $P_i=P_{i}^\text{th}$. Table 1 summarises the parameters used to simulate each of the 16 energy spectra. Columns two and three list the tunnel couplings in $\mu\text{eV}$ between the spin manifolds present in the energy spectrum at the given charge occupation (see Appendix~\ref{Sec6}), whereas columns four and five list the energy splittings in $\mu\text{eV}$ at large negative (positive) detuning $\delta_\text{N}$ ($\delta_\text{P}$). Note that values of $\delta_\text{N}= \delta_\text{P}>120~\mu\text{eV}$ simply indicate that the splitting is larger than $120~\mu\text{eV}$ and hence renders the high-spin anti-crossing unobservable within the measured and simulated range of magnetic fields. Finally, the last column denotes the relaxation regime simulated, i.e. whether tunneling capacitance is included or not. We use a temperature of $T=80$~mK for all simulations and $\delta_\text{N(P)}=18~\mu\text{eV}$ for all splittings that correspond to the T$_3$-T$_4$ separation $\delta$.

\begin{figure*}[h]
	\centering
	\includegraphics[width=1\textwidth]{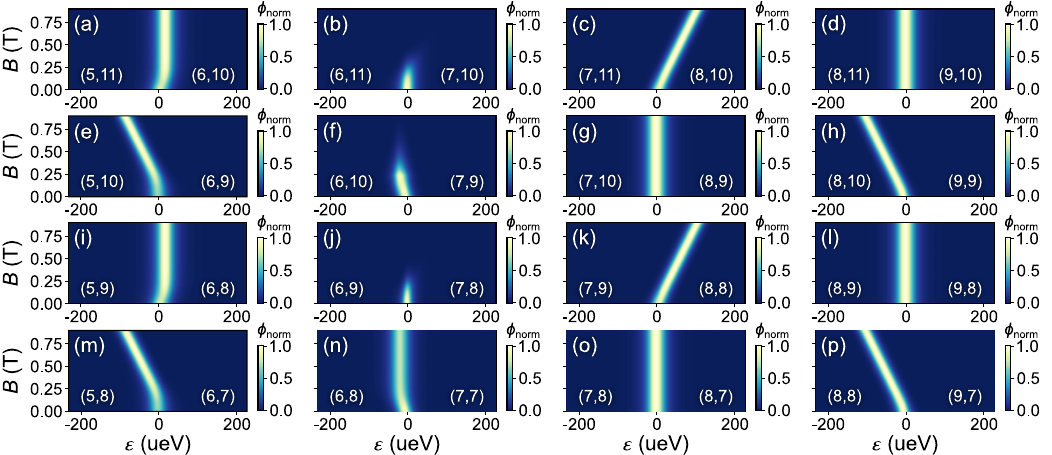}
	\caption{Magnetospectroscopy simulations of the 16 studied ICTs. \textbf{(a)-(p)} Qualitative simulation of the normalised phase shift of each ICT visible in Fig.~\ref{fig1}(b) and their energy spectra in Figs.~\ref{sfig4} and \ref{sfig5} as a function of $B$ using the parameters in Table 1. The numbers in parentheses indicate the electron occupancy of the DQD on either side of the ICT.}
	\label{sfig6}
\end{figure*}

Based on the proportionality $\phi \propto \Delta C_\text{pm}$, we use Eqs. \eqref{Cpm2} and \eqref{polarization2} and the parameters in Table 1 to qualitatively simulate the normalised phase shift $\phi_\text{norm}$ and magnetospectra for the energy spectra sketched in Figs.~\ref{sfig4} and \ref{sfig5}. The simulation results are shown in Fig.~\ref{sfig6}. For the simulation of Fig.~\ref{sfig6}(f), we introduce a suppression of the phase shift signal for $B>0.2$~T to account for Landau-Zener transitions which cause the change in QD occupation probability to approach zero. Overall, the simulations represent the measured magnetospectra in Fig.~\ref{fig3} well.

	\onecolumngrid
	
	\begin{center}
		\begin{table}[h]
			\label{sim_params}
			\begin{tabular}{@{}cccccc@{}}
				\toprule
				\multicolumn{1}{|c|}{Panel} & \multicolumn{1}{c|}{$2t_\text{S}$ or $2t_\text{D}$ ($\mu$eV)} & \multicolumn{1}{c|}{$2t_\text{T}$ or $2t_\text{q}$  ($\mu$eV)} & \multicolumn{1}{c|}{$\delta_\text{N}$  ($\mu$eV)}         & \multicolumn{1}{c|}{$\delta_\text{P}$  ($\mu$eV)}   & \multicolumn{1}{c|}{Relaxation regime} \\ \midrule
				\multicolumn{1}{|c|}{a} & \multicolumn{1}{c|}{20} & \multicolumn{1}{c|}{20} & \multicolumn{1}{c|}{0} & \multicolumn{1}{c|}{18}  & \multicolumn{1}{c|}{Slow}                \\
				\multicolumn{1}{|c|}{b} & \multicolumn{1}{c|}{15} & \multicolumn{1}{c|}{20} & \multicolumn{1}{c|}{18}  & \multicolumn{1}{c|}{\textgreater{}120}& \multicolumn{1}{c|}{Slow}   \\
				\multicolumn{1}{|c|}{c} & \multicolumn{1}{c|}{10} & \multicolumn{1}{c|}{20} & \multicolumn{1}{c|}{0}  & \multicolumn{1}{c|}{\textgreater{}120} & \multicolumn{1}{c|}{Fast}  \\
				\multicolumn{1}{|c|}{d} & \multicolumn{1}{c|}{20} & \multicolumn{1}{c|}{20} & \multicolumn{1}{c|}{\textgreater{}120} & \multicolumn{1}{c|}{\textgreater{}120} & \multicolumn{1}{c|}{Fast}  \\
				\multicolumn{1}{|c|}{e} & \multicolumn{1}{c|}{20} & \multicolumn{1}{c|}{20} & \multicolumn{1}{c|}{\textgreater{}120} & \multicolumn{1}{c|}{18}    & \multicolumn{1}{c|}{Fast}  \\
				\multicolumn{1}{|c|}{f} & \multicolumn{1}{c|}{10} & \multicolumn{1}{c|}{3}  & \multicolumn{1}{c|}{18}   & \multicolumn{1}{c|}{0}      & \multicolumn{1}{c|}{Fast}  \\
				\multicolumn{1}{|c|}{g} & \multicolumn{1}{c|}{20} & \multicolumn{1}{c|}{20} & \multicolumn{1}{c|}{\textgreater{}120} & \multicolumn{1}{c|}{\textgreater{}120} & \multicolumn{1}{c|}{Fast}  \\
				\multicolumn{1}{|c|}{h} & \multicolumn{1}{c|}{10} & \multicolumn{1}{c|}{20} & \multicolumn{1}{c|}{\textgreater{}120} & \multicolumn{1}{c|}{0}  & \multicolumn{1}{c|}{Fast}   \\
				\multicolumn{1}{|c|}{i} & \multicolumn{1}{c|}{20} & \multicolumn{1}{c|}{20} & \multicolumn{1}{c|}{0} & \multicolumn{1}{c|}{18}  & \multicolumn{1}{c|}{Slow}  \\
				\multicolumn{1}{|c|}{j} & \multicolumn{1}{c|}{7}  & \multicolumn{1}{c|}{20} & \multicolumn{1}{c|}{18}     & \multicolumn{1}{c|}{\textgreater{}120} & \multicolumn{1}{c|}{Slow}  \\
				\multicolumn{1}{|c|}{k} & \multicolumn{1}{c|}{10} & \multicolumn{1}{c|}{20} & \multicolumn{1}{c|}{0}   & \multicolumn{1}{c|}{\textgreater{}120}& \multicolumn{1}{c|}{Fast}   \\
				\multicolumn{1}{|c|}{l} & \multicolumn{1}{c|}{20} & \multicolumn{1}{c|}{20} & \multicolumn{1}{c|}{\textgreater{}120} & \multicolumn{1}{c|}{\textgreater{}120}& \multicolumn{1}{c|}{Fast}   \\
				\multicolumn{1}{|c|}{m} & \multicolumn{1}{c|}{20} & \multicolumn{1}{c|}{20} & \multicolumn{1}{c|}{\textgreater{}120} & \multicolumn{1}{c|}{18}     & \multicolumn{1}{c|}{Fast}   \\
				\multicolumn{1}{|c|}{n} & \multicolumn{1}{c|}{10} & \multicolumn{1}{c|}{20} & \multicolumn{1}{c|}{18}   & \multicolumn{1}{c|}{0}    & \multicolumn{1}{c|}{Fast}   \\
				\multicolumn{1}{|c|}{o} & \multicolumn{1}{c|}{20} & \multicolumn{1}{c|}{20} & \multicolumn{1}{c|}{\textgreater{}120} & \multicolumn{1}{c|}{\textgreater{}120}& \multicolumn{1}{c|}{Fast}   \\
				\multicolumn{1}{|c|}{p} & \multicolumn{1}{c|}{10} & \multicolumn{1}{c|}{20} & \multicolumn{1}{c|}{\textgreater{}120} & \multicolumn{1}{c|}{0} & \multicolumn{1}{c|}{Fast}  \\ \bottomrule
			\end{tabular}
			\caption{Parameters used to perform qualitative magnetospectroscopy simulations shown in Fig.~\ref{sfig6} for the energy levels shown in Figs.~\ref{sfig4} and \ref{sfig5}. Columns two through five list the magnitude of the needed energy spectrum parameters, and the last column indicates whether the simulations consider the slow or fast relaxation regime. All simulations use $T=80$~mK.}
		\end{table}
	\end{center}

\clearpage

\twocolumngrid
\bibliography{nonreciprocalPSB}
\end{document}